\documentclass[onecolumn,superscriptaddress,floatfix,showpacs]{revtex4-2}

\usepackage{tabularx}

\usepackage[utf8]{inputenc}
\usepackage[T1]{fontenc}     
\usepackage[british]{babel}  
\usepackage[sc,osf]{mathpazo}
\usepackage[scaled=0.86]{berasans}  
\usepackage[colorlinks=true, linkcolor=blue,citecolor=blue, urlcolor=blue]{hyperref}  
\usepackage{mathtools}
\usepackage{lipsum}
\usepackage{graphicx} 
\usepackage{subfig}
\usepackage[babel]{microtype}  
\usepackage{amsmath,amssymb,amsthm,bm,amsfonts,mathrsfs,bbm} 

\usepackage{xspace}  
\usepackage{pgfplots}
\usepackage{xcolor,colortbl}
\def\ba{\begin{equation}}
	\def\ea{\end{equation}}
\def\bea{\begin{eqnarray}}
	\def\eea{\end{eqnarray}}
\def\ben{\begin{equation*}}
	\def\een{\end{equation*}}
\def\bean{\begin{eqnarray*}}
	\def\eean{\end{eqnarray*}}
\def\bma{\begin{mathletters}}
	\def\ema{\end{mathletters}}
\def\bi{\begin{itemize}}
	\def\ei{\end{itemize}}

\newcommand{\be}{\begin{equation}}
	\newcommand{\ee}{\end{equation}}

\renewcommand{\AA}{\ensuremath{\mathcal{A}}}

\newcommand{\kommentar}[1]{}

\newcommand{\forget}[1]{}

\newtheorem{theorem}{Theorem}

\newtheorem{corollary}{Corollary}[theorem]

\newtheorem{observation}{Observation}
\begin{document}
	\title{On fully entangled fraction and quantum conditional entropies for states with maximally mixed marginals}
	
	\author{Komal Kumar}
	\email{p20210063@hyderabad.bits-pilani.ac.in}
	\affiliation{Department of Mathematics, Birla Institute of Technology and Science Pilani, Hyderabad Campus,Telangana-500078, India}
	\author{Indranil Chakrabarty}
	\email{indranil.chakrabarty@iiit.ac.in}
	\affiliation{Centre for Quantum Science and Technology and Center for Security, Theory and Algorithmic Research, International Institute of Information Technology, Hyderabad, Gachibowli, Telangana-500032, India}
	\author{Nirman Ganguly}
	\email{nirmanganguly@hyderabad.bits-pilani.ac.in}
	\affiliation{Department of Mathematics, Birla Institute of Technology and Science Pilani, Hyderabad Campus,Telangana-500078, India}
	
	\begin{abstract}

        The fully entangled fraction (FEF) measures the proximity of a quantum state to maximally entangled states. FEF $>\frac{1}{d}$, in $d \otimes d$ systems is a significant benchmark for various quantum information processing protocols including teleportation. Quantum conditional entropy (QCE) on the other hand is a measure of correlation in quantum systems. Conditional entropies for quantum systems can be negative, marking a departure from conventional classical systems. The negativity of quantum conditional entropies plays a decisive role in tasks like state merging and dense coding. In the present work, we investigate the relation of these two important yardsticks. Our probe is mainly done in the ambit of states with maximally mixed marginals, with a few illustrations from other classes of quantum states. We start our study in two qubit systems, where 
        for the Werner states, we obtain lower bounds to its FEF when the conditional Rényi $\alpha-$entropy is negative. We then obtain relations between FEF and QCE for two qubit Weyl states. Moving on to two qudit states we find a necessary and sufficient condition based on FEF, for the isotropic state to have negative conditional entropy. In two qudit systems the relation between FEF and QCE is probed for the rank deficient and generalized Bell diagonal states. FEF is intricately linked with $k$- copy nonlocality and $k$- copy steerability. The relations between FEF and QCE facilitates to find conditions for $k$- copy nonlocality and $k$- copy steerability based on QCE. We obtain such conditions for certain classes of states in two qubits and two qudits. Applications of the relations obtained are provided in the context of work extraction, faithful entanglement and entropic uncertainty relations. 
  
  \end{abstract}
	\date{\today}
	\maketitle

		\section{Introduction}
		The study on quantum correlations is an area of intense research due to its implications in various information processing tasks. There are various forms of  quantum correlations. Entanglement \cite{horodecki2009quantum} is one of such correlations that has been widely discussed followed by other forms of quantum correlations like quantum discord \cite{ollivier2001quantum,chakrabarty2011quantum,modi2010unified}. In addition, quantum conditional entropies \cite{cerf1999quantum} (QCE) and fully entangled fraction (FEF) \cite{horodecki1999general} are efficient quantifiers of quantum correlation and can be indicators of resourcefulness of a quantum state in quantum information processing tasks.
		
      Entropies \cite{wehrl1978general} which measure randomness and uncertainty is standardized by  the Shannon entropy \cite{shannon1948mathematical} of a random variable with a given probability distribution.  The idea of Shannon entropy can be generalized to other  various types of entropies, namely, Rényi $\alpha$- entropies \cite{renyi1961measures}, Tsallis $\alpha$- entropies \cite{tsallis1988possible}. These entropies and their quantum counterparts \cite{muller2013quantum,von2018mathematical}, are used in classical and quantum scenarios.  
      
      The quantum version of these entropies has been widely used to establish various types of inequalities, namely, entropic Bell inequalities \cite{cerf1997entropic}, entropic steering inequalities \cite{schneeloch2013einstein,costa2018entropic} and entropic uncertainty relations \cite{wehner2010entropic}, that have applications in quantum cryptography \cite{gisin2002quantum}. In quantum scenarios, conditional entropies can be negative \cite{cerf1997negative} whereas it is always non-negative in the classical realm. The negativity of quantum conditional entropies provides an advantage in quantum information processing tasks. Quantum protocols, namely, super-dense coding \cite{bennett1992communication,bruss2004distributed,prabhu2013exclusion} and quantum state merging \cite{horodecki2005partial,horodecki2007quantum} provides an operational interpretation to the negativity of conditional entropies. Recently, in \cite{vempati2021witnessing,vempati2022unital,brandsen2021quantum,patro2017non}, it has been observed that negativity of conditional von Neumann entropy acts as a quantum resource. A recent work  \cite{gour2022inevitability}, posits that quantum conditional entropies are inherently negative in any axiomatic characterization.
  
		Sometimes, the mere presence of entanglement is not enough. Certain quantum protocols fail in quantum information processing as not every entangled state acts as a resource. A yardstick for the efficacy of an entangled state is provided by its fully entangled fraction \cite{horodecki1999general}. The fully entangled fraction of an entangled state beyond some threshold value acts as a resource \cite{horodecki1999general}, and has operational interpretation in context of quantum teleportation \cite{bennett1993teleporting}, entanglement swapping \cite{zukowski1993event} and quantum cryptography \cite{gisin2002quantum}. The explicit calculation of FEF for quantum states is in general hard. In \cite{li2008upper,huang2016upper}, the upper bound of fully entangled fraction has been discussed, and also the estimation of fully entangled fraction has been obtained in \cite{rui2010estimation}. A plethora of research works have been carried to further study the properties of fully entangled fraction in relation to various aspects of quantum information processing \cite{zhao2015maximally,Zhao_2010,grondalski2002fully,patro2022absolute,ganguly2011entanglement,cavalcanti2013all}. 
  
        Since some threshold values of FEF and and negativity of QCE are significant to various quantum information processing tasks, the present work investigates the relation between these two significant quantifiers and the implications these relations carry pertaining to quantum information processing protocols. The motivation for our contribution can thus broadly be stated as follows:
        We identify the common class of quantum states which are useful for some quantum protocols. The quantum protocols considered here are quantum teleportation and superdense coding and are characterized by FEF$>\frac{1}{d}$ and negative conditional von Neumann entropy, respectively. So, in this perspective, our work characterizes the resource useful for teleportation based on its negative entropic measure.  In the context of quantum nonlocality, a nonlocal resource provides the quantum advantage in the security of quantum cryptography protocols \cite{acin2007device,acin2006bell}, in communication complexity protocols \cite{buhrman2010nonlocality}, and in the generation of trusted random numbers \cite{pironio2010random}. From this perspective, our work characterizes the  $k-$copy nonlocality and $k-$copy steerability based on its negative entropic measure.
        In some works in the recent past, the relation between quantum nonlocality and negative QCE has been investigated \cite{friis2017geometry,kumar2023quantum}. Particularly in \cite{friis2017geometry}, the authors investigated the relation between Bell-CHSH inequality \cite{clauser1969proposed} and QCE, whereas some of us investigated the relation between the CJWR steering inequality \cite{cavalcanti2009experimental} and QCE in \cite{kumar2023quantum}. However, to the best of our knowledge, these studies have not been done in the context of $k-$copy nonlocality \cite{cavalcanti2013all,palazuelos2012superactivation} and $k-$copy steerability \cite{quintino2016superactivation}.

		In the present work, we begin our investigation with  $2 \otimes 2$ system, wherein we probe the relation of quantum conditional entropies and fully entangled fraction for the Weyl state (states with maximally mixed marginals). In particular, for Werner state, we show that the negativity of conditional von Neumann entropy is a sufficient condition for fully entangled fraction strictly greater than half, and further, we obtain upper bounds to the conditional von Neumann entropy whenever it is useful for teleportation. We observe that fully entangled fraction of two-qubit Werner states has a lower bound in terms of eigenvalues whenever it possesses negative conditional Rényi $\alpha-$entropies. For the two-qubit Weyl state, we show that the negativity of conditional Rényi 2-entropies is a sufficient condition for fully entangled fraction strictly greater than half, and the upper bounds of conditional Rényi 2-entropies are provided if it is useful for teleportation. In the $d \otimes d$ system, for isotropic state \cite{horodecki1999reduction}, we obtain the lower bound of fully entangled fraction whenever its conditional von Neumann entropy is negative. Lower bounds of fully entangled fraction of two-qudit isotropic states is also provided, which guarantees the negativity of conditional Rényi 2-entropies and vice versa. Also, we discuss the relation between fully entangled fraction and conditional entropies for two qudit Werner states, rank deficient states and generalized Bell diagonal states. We then investigate the relation between $k-$copy steerability and conditional entropies of the isotropic state. We find that a two-qubit isotropic state is $k-$copy steerable if conditional von Neumann entropy is negative, and the same holds for conditional Rényi 2-entropies. Moving from $k-$copy steerability to $k-$copy nonlocality, the negativity of conditional Rényi 2-entropy is a sufficient condition for $k-$copy nonlocality of the two-qubit Weyl state. Some two-qubit non-Weyl states show the same behavior under some restriction. We then consider the mixture of arbitrary pure state and maximally mixed noise \cite{cavalcanti2013all} for dimension $d=3,4,5$ and observe that it is $k-$copy nonlocal whenever they possess negative conditional Rényi 2-entropy and conditional von Neumann entropy. Furthermore, as a corollary to the relations obtained between FEF and QCE, we present an application in the context of work extraction  \cite{rio2011thermodynamic}, faithful entanglement \cite{weilenmann2020entanglement,riccardi2021exploring,guhne2021geometry}, and entropic uncertainty relation \cite{robertson1929uncertainty,kraus1987complementary,maassen1988generalized,berta2010uncertainty}.
		
		This manuscript is arranged in the following manner: In sec. \ref{II}, we discuss the notation and preliminaries required for our work. We establish the relation of conditional entropies and fully entangled fraction for $2 \otimes 2$ system and $d \otimes d$ system in sec. \ref{III} and \ref{IV}, respectively. In sec. \ref{V}, we study  $k-$copy steerability and conditional entropies, and then discuss the $k-$copy nonlocality and conditional entropies in sec \ref{VI}. Further, we explore the application of fully entangled fraction in sec. \ref{VII}. In sec. \ref{VIII}, we give our concluding remarks emphasizing on some future work.		
		\section{Notation and Preliminaries} \label{II}
		In this section, we fix the notations, and briefly discuss essential concepts that are important for subsequent study. In this work, we deal with finite-dimensional Hilbert spaces. A quantum system denoted by $ X(=A, B) $ is described by finite-dimensional Hilbert space $\mathbb{H}_{X}$. Our work is mainly based on bipartite quantum system $AB = A \otimes B$ that is characterized by Hilbert space $\mathbb{H}_{AB}=\mathbb{H}_{A} \otimes \mathbb{H}_{B}$. The Hilbert-Schmidt space $ \mathfrak{B}(\mathbb{H}_{AB}) $ is the space of bounded linear operators acting on $\mathbb{H}_{AB}$.  Quantum states are described by density operators $ \rho \in  \mathfrak{B}(\mathbb{H}_{AB})$ such that it is positive semi-definite ($ \rho \ge 0 $), Hermitian (which also follows from positivity) with unit trace. The von Neumann entropy, Rényi $ \alpha- $ entropy and Tsallis $ \alpha- $ entropy of the quantum state  are denoted by $ S(\cdot) , S_\alpha(\cdot), S^T_\alpha(\cdot) $, respectively. $\text{FEF}(\cdot)$ denotes the fully entangled fraction of the quantum state.

		\subsection{Bloch-Fano Decomposition of density matrices}
		
		In composite quantum systems (in particular bipartite systems), quantum state is described by density matrices, which can be represented as \cite{friis2017geometry}
			\begin{eqnarray}
            \rho_{d_A \otimes d_B} =\frac{1}{d_A d_B} [\mathbb{I}_A \otimes \mathbb{I}_B + \sum_{i=1}^{d_A^2 - 1} a_i g_i^A \otimes \mathbb{I}_B + \sum_{j=1}^{d_B^2 - 1} b_j  \mathbb{I}_A \otimes g_j^B + \sum_{i=1}^{d_A^2 - 1} \sum_{j=1}^{d_B^2 - 1} t_{ij} g_i^A \otimes g_j^B], \label{e1}
			\end{eqnarray}
		where $ \text{dim}~ \mathbb{H}_{A} = d_A  $ and $ \text{dim}~ \mathbb{H}_{B} = d_B  $. The hermitian operators $ g_i^k $ for $ k=A,B $ are generalizations of the Pauli matrices, i.e., they are orthogonal $ Tr[g_i^k g_j^k] = 2 \delta_{ij} $ and traceless, $ Tr[g_i^k] =0  $ and for single qubit systems they coincide with the Pauli matrices. The coefficients $ a_i, b_j \in \mathbb{R} $ are the components of the generalized Bloch vectors $ \vec{a}, \vec{b} $ of the subsystems $ A,B $, respectively. The real coefficients $ t_{ij} $ are the components of the correlation tensor. For two-qubit systems the density matrices can be represented as, 
		\begin{eqnarray}
			\rho_{2 \otimes 2}=\frac{1}{4} [\mathbb{I}_{2}\times\mathbb{I}_2+\vec{\mathfrak{a}}.\vec{\sigma}\otimes \mathbb{I}_2+\mathbb{I}_2\otimes \vec{\mathfrak{b}}.\vec{\sigma} + \sum_{i=1}^{3} \sum_{j=1}^{3} t_{ij} \sigma_i^A \otimes \sigma_j^B ].\label{e2}
		\end{eqnarray} 
		
		An interesting class of states is the \textit{locally maximally mixed states or Weyl states} which in the two-qudit systems (up to local unitaries) is given by, 
		
		\begin{equation}
			\rho^{weyl}_{d} = \frac{1}{d^2} [\mathbb{I}_A \otimes \mathbb{I}_B + \sum_{i=1}^{d^2 - 1} w_{i} g_i^A \otimes g_i^B].\label{e3}
		\end{equation}
		The reduced marginals of such states are maximally mixed, i.e., $ \text{Tr}_A [\rho^{weyl}_{d}] =\frac{\mathbb{I}_A}{d}  $ and $  \text{Tr}_B [\rho^{weyl}_{d}] =\frac{\mathbb{I}_B}{d}  .$
		\subsection{Quantum entropies}
		In this subsection we give the definition of the list of entropies that we have used in our work. The von Neumann entropy of a quantum state $\rho_{AB} \in  \mathfrak{B}(\mathbb{H}_{AB}) $  is defined as,
		\begin{equation}
			S(\rho_{AB}) = -\text{Tr} \left[\rho_{AB} \log_2 \rho_{AB}\right], \label{e4}
		\end{equation}
		where the logarithms are taken to the base 2.
		The von Neumann entropy has a special association with the eigenvalues of the density matrix, i.e., it is a function of the eigenvalues. The corresponding conditional von Neumann entropy (CVNE) is given by $ S(A|B)=S(\rho_{AB})-S(\rho_B) $.\\
		
		\noindent The Rényi $\alpha$-entropy is given by
		\begin{align}
			S_{\alpha}(\rho_{AB})= \frac{1}{1-\alpha} \log_{2}\left[Tr(\rho_{AB}^{\alpha})\right],~\alpha >0,~~\alpha \neq 1. \label{e5}
		\end{align}
		The von Neumann entropy is the limiting case of the Rényi entropy as 
		$\alpha\rightarrow 1$. The corresponding conditional Rényi $\alpha$- entropy (CRAE) is given by $ S_\alpha(A|B)=S_\alpha(\rho_{AB})-S_\alpha(\rho_B) $.\\
		
		\noindent Similarly, Tsallis $\alpha$-entropy is given by 
		\begin{align}
			S^T_{\alpha}(\rho_{AB})= \frac{1}{1-\alpha} [Tr(\rho_{AB}^{\alpha})-1],~\alpha >0,~\alpha \neq 1. \label{e6}
		\end{align}
		The conditional Tsallis $\alpha$-entropy (CTAE) is given by \cite{vollbrecht2002conditional}, 
		\begin{align*}
			S^T_{\alpha}(A|B)=\frac{Tr(\rho_B^\alpha)-Tr(\rho_{AB}^\alpha)}{(\alpha-1)Tr(\rho_B^\alpha)}.
		\end{align*}
	\subsection{Fully entangled fraction}
	 In a bipartite quantum system, for any quantum state $\rho_{AB} \in  \mathfrak{B}(\mathbb{H}_{AB}) $, where $\mathbb{H}_{AB}=\mathbb{H}_{A}\otimes \mathbb{H}_{B}$, $\text{dim}~ \mathbb{H}_{A} = \text{dim}~ \mathbb{H}_{B} = d$, the fully entangled fraction $\text{FEF}(\rho_{AB})$ is defined as the maximal overlap of quantum state $\rho_{AB}$ with maximally entangled pure state. In other words, we can also say that it is a measure of proximity of quantum state with maximally entangled pure state. The fully entangled fraction of state $\rho$ is given as 
  \begin{align}
      \text{FEF}(\rho_{AB}) = \text{max}_{|\phi\rangle\in S}\langle\phi|\rho_{AB}|\phi\rangle, \label{e7a}
  \end{align}
  where $S$ is the set of all maximally entangled pure states. In other way, Equation(\ref{e7a}) can be expressed as
   \cite{horodecki1999general} 
	 \begin{align}
   	 	\text{FEF}(\rho_{AB}) = \text{max}_{U}\langle\psi_{d}^{+}|U\otimes I\rho U^{\dagger}\otimes I |\psi_{d}^{+}\rangle, \label{e7}
     \end{align}
    where $|\psi_{d}^{+}\rangle=\frac{1}{\sqrt{d}}\sum_{i=1}|ii\rangle$ maximally entangled pure state, $ U$ is unitary transformation and $I$ is identity transformation. In $2\otimes2$ system, fully entangled fraction of any quantum state $\rho_{AB}$ is given as \cite{horodecki1997inseparable} 
    \begin{align}
    	\text{FEF}(\rho_{AB}) = \frac{1}{4}\left[1+ N(\rho_{AB})\right], \label{e8}
    \end{align}
     where $N(\rho)=\text{Tr}|T|$; $|T|=\sqrt{T^{\dagger}T}$ and $T=[t_{ij}]$ is the correlation tensor of the general two qubit density matrix $\rho_{AB}$.
     
     \subsection{Quantum nonlocality}
     \subsubsection{{\textbf{k-copy nonlocality}}}
      Consider two parties, Alice and Bob sharing an entangled state $\rho_{AB}$. They perform measurements on their respective subsystem and obtain corresponding outcomes. Quantum measurements performed by Alice and Bob are characterized by set of positive operator $\{A_{a|x}\}$ \& $\{B_{b|y}\}$ such that $ \sum_{a,x}A_{a|x}=I $ \& $ \sum_{b,y}B_{b|y}=I $, respectively. Here, $a$ \& $b$ are their respective outcomes corresponding to their choice of measurements labeled by $x$ \& $y$. The resulting joint probabilities of outcomes are given by 
      \begin{align}
      	p(a,b|x,y)= \text{Tr}[(A_{a|x}\otimes B_{b|y})\rho_{AB}]. \label{e9}
      \end{align}
       For some hidden variable $\lambda$ with probability distribution $q(\lambda)$, if all the above resulting joint probabilities of outcomes can be expressed as 
       \begin{align}
       	p(a,b|x,y) = \int_{\lambda}p_{A}(a|x,\lambda)p_{B}(b|y,\lambda)q(\lambda)d\lambda, \label{e10}
       \end{align}
       where $p_{A}(a|x,\lambda)$ \& $p_{B}(b|y,\lambda)$ represent the local resulting statistics of Alice and Bob, respectively, then the shared entangled state $\rho_{AB}$ is said to be local or admit local hidden variable (LHV) model. If $p(a,b|x,y)$ does not admit decomposition of the form of Equation (\ref{e10}), then entangled state $\rho_{AB}$ is nonlocal, and hence violates some chosen Bell's inequality for some measurements $\{A_{a|x}\}$ \& $\{B_{b|y}\}$ \cite{bell1964einstein,brunner2014bell}.
       
       In standard Bell scenario, separated parties, Alice and Bob, perform measurements on single copy of entangled state $\rho_{AB}$ in each round of test. Several rounds are required to test the Bell's inequality. In case of $k-$ copy Bell scenario, parties perform local joint measurements on several copies of entangled state, namely $k-$copy (i.e., $\rho_{AB}^{\otimes k}$ known as global bipartite entangled state) \cite{quintino2016superactivation}. Now, we can have the case where initially the entangled state is local or has LHV model, but after performing local collective measurements on $k-$copy of entangled state it becomes Bell nonlocal and violates some chosen Bell's inequality. Then, we say that quantum state $\rho_{AB}$ is $k-$copy nonlocal for some $k$.
     \subsubsection{{\textbf{k-copy steering}}}
     Quantum steering is a different notion of nonlocality. In standard quantum steering scenario, there are two spatially separated parties Alice and Bob sharing an entangled state $\rho_{AB}$. Here, only Alice performs some measurements on her part of system. Measurements are characterized by positive operator $\{A_{a|x}\}$ such that $ \sum_{a,x}A_{a|x}=I $, and choice of measurements are labeled by $x$, and $a$ are the corresponding outcomes. After performing measurements, Alice remotely steers the state of Bob's system to unnormalized conditional state $\rho_{a|x}$ (is known as quantum steering assemblage) for each measurement $A_{a|x}$. The resulting quantum steering assemblage is given by 
     \begin{align}
     	\rho_{a|x}= \text{Tr}_{A}[(A_{a|x}\otimes I)\rho_{AB}], \label{e11}
     \end{align}
 where $\text{Tr}_{A}$ is partial trace on Alice's system. Now, if the above resulting conditional state (steering assemblage) on Bob's side can be expressed as
 \begin{align}
 	\rho_{a|x}=\int_{\lambda}\sigma_{\lambda}p_{A}(a|x,\lambda)q(\lambda)d\lambda, \label{e12}
 \end{align}
where $\sigma_{\lambda}$ is hidden state on Bob's side corresponding to hidden variable $\lambda$ with probability distribution $q(\lambda)$ and $p_{A}(a|x,\lambda)$ represents the local resulting statistics of Alice. Then, the quantum state $\rho_{AB}$ is said to be unsteerable or has local hidden state (LHS) model. If $\rho_{a|x}$ does not admit decomposition of the form of Equation (\ref{e12}), then entangled state $\rho_{AB}$ is steerable from Alice to Bob, and hence violates some chosen steering inequality for some measurement $A_{a|x}$.

In standard steering scenario, Alice performs some measurements on single copy of entangled state $\rho_{AB}$ in each round of test. Several rounds are required to test the steering inequality \cite{cavalcanti2009experimental}. In case of $k-$copy steering scenario, Alice performs local joint measurements on several copies of entangled state, namely $k-$copy (i.e., $\rho_{AB}^{\otimes k}$ known as global bipartite entangled state). Now, we can have the case where initially the entangled state is unsteerable or has LHS model, but after performing local collective measurements on $k-$copy of entangled state it becomes steerable and violates some chosen steering inequality \cite{cavalcanti2009experimental}. Then, we say that quantum state $\rho_{AB}$ is $k-$copy steerable for some $k$.

Having fixed the notations and basic definitions, we proceed to our study in the next section.
 
		\section{Conditional Entropies and Fully Entangled fraction for two-qubit states} \label{III}

  In this section we consider the relationship of the entropic functions like conditional von Neumann entropies and $\alpha$-conditional entropies with fully entangled fraction for different two qubit states like Werner state and Weyl state. Though our results are dependent on the nature of the state, however it gives us a perspective on how the quantifiers of entanglement are connected with entropic measures. As noted earlier, negativity of conditional von Neumann entropy is a benchmark for superdense coding and state merging with quantum advantage. Similarly $\text{FEF} > 1/d$ for $d \otimes d$ systems decides the efficacy of quantum states in the standard teleportation protocol. Thus the relations between FEF and QCE that we obtain in this and the subsequent sections  identify a common class of quantum states which are useful for both teleportation and superdense coding thus opening up the opportunity of using the same resource for multiple tasks. 
		\subsection{Werner States}
  The two-qubit Werner state  in the Pauli basis $\{\sigma_0,\sigma_1,\sigma_2,\sigma_3\}$ is given by,
  \begin{equation}
   \rho^{wer}_{2}(p)= \frac{1}{4}[\mathbb{I}_2 \otimes \mathbb{I}_2-\sum_{i=1}^{3}p(\sigma_{i} \otimes \sigma_{i})],   
  \end{equation}
where $\mathbb{I}_2$ is the $2 \times 2$ identity matrix and $p$ is the visibility parameter.   
		\subsubsection{{\textbf{Relation with conditional von Neumann entropy}}}\label{vnew}
  The conditional von Neumann entropy (CVNE) for the Werner state is given by, 
				\begin{eqnarray}
				S(A|B)_{\rho^{wer}_{2}(p)}=-3\left( \frac{1-p}{4}\right) \log_{2}\left(\frac{1-p}{4}\right)-\left(\frac{1+3p}{4}\right)\log_{2}\left(\frac{1+3p}{4}\right)-1 .\label{e13}
			\end{eqnarray}
		CVNE is negative for $ p >  0.747614 $. The fully entangled fraction of Werner state is given by $\text{FEF}(\rho^{wer}_2)=\frac{1}{4}[1+N(\rho^{wer}_2)]$, where $N(\rho^{wer}_2)=Tr|T|$ and $T$ is correlation matrix. So, $\text{FEF}(\rho^{wer}_2)>\frac{1}{2}\Leftrightarrow p>\frac{1}{3}$. Thus, if the Werner state has negative conditional von Neumann entropy (CVNE) then it is useful for teleportation. 

         We now derive bounds to the FEF and QCE based on the relations between them.   
		\begin{theorem}\label{t1}
			If fully entangled fraction of Werner state is strictly greater than half i.e., $\text{FEF}(\rho^{wer}_2)>\frac{1}{2}$, then conditional von Neumann entropy \text{(CVNE)} is upper bounded by $3\delta\log_{2}(1/2\delta)$ i.e., 
		\begin{align*}
			\text{FEF}(\rho^{wer}_2)>\frac{1}{2} \Rightarrow \text{CVNE}(\rho^{wer}_2) < 3\delta\log_{2}(1/2\delta),
		\end{align*}
		where $\delta=\text{min}\{\delta_1,\delta_2\}$, $\delta_1=\frac{1-p}{4}$
        ~~\text{and}~~$\delta_2= \frac{1+3p}{4}$ are eigenvalues of  $\rho^{wer}_2$ with multiplicity $3$ and $1$, respectively.
		\end{theorem}
	       \begin{eqnarray}
            &&\text{{\bf{Proof:}} Let the fully entangled fraction } \text{FEF}(\rho^{wer}_2) >\frac{1}{2}{}\nonumber\\&&
			\Rightarrow \frac{1+3p}{4}  > \frac{1}{2}
			\Rightarrow \frac{4}{1+3p}  < 2
			\Rightarrow \log_{2}\left[\frac{4}{1+3p}\right]  < 1 {}\nonumber\\&&
			\Rightarrow \left(\frac{1+3p}{4}\right) \log_{2}\left[\frac{4}{1+3p}\right]  < \frac{1+3p}{4}{}\nonumber\\&&
			 \Rightarrow-3\left( \frac{1-p}{4}\right) \log_{2}\left(\frac{1-p}{4}\right)-\left(\frac{1+3p}{4}\right)\log_{2}\left(\frac{1+3p}{4}\right)-1 {}\nonumber\\&& < \frac{1+3p}{4} - 3\left(\frac{1-p}{4}\right)\log_{2}\left(\frac{1-p}{4}\right) - 1{}\nonumber\\&&
			\Rightarrow \text{CVNE}(\rho^{wer}_2)  < 3 \delta \log_{2}(1/2\delta).{}\nonumber
		\end{eqnarray}
This result can also be interpreted as one that gives the bound on the superdense coding capacity of the states that are useful for teleportation.
    \subsubsection{{\textbf{Relation with other $ \alpha- $ conditional entropies}}}
		Conditional Rényi $\alpha$- entropy (CRAE) of two-qubit Werner state $\rho^{wer}_{2}(p)$ is given by
		\begin{eqnarray}
			S_{\alpha}(A|B)_{\rho^{wer}_{2}(p)}=&& \frac{1}{1-\alpha}\log_{2}[\frac{(\frac{1+3p}{4})^{\alpha}+3 \left(\frac{1-p}{4}\right)^{\alpha}}{2(\frac{1}{2})^{\alpha}}]~;~~~\alpha > 1 \label{e14}
		\end{eqnarray}
		In the following, we obtain lower bounds for fully entangled fraction whenever its conditional Rényi $\alpha-$entropy is negative. Since  $ S^T_{\alpha}(A|B) \ge 0 \Leftrightarrow S_{\alpha}(A|B) \ge 0  $, we arrive at the same conclusion for the conditional Tsallis $\alpha-$entropy.\\
		\begin{theorem}\label{t2}
			If conditional Rényi $\alpha-$entropy (CRAE) of two-qubit Werner state is negative, then the fully entangled fraction is lower bounded by $[\frac{1}{2^{\alpha-1}}-3\delta^{\alpha}]^{1/\alpha}$ i.e.,
		\begin{align*}
		\text{CRAE}(\rho^{wer}_2) < 0 \Rightarrow FEF(\rho^{wer}_2)>[\frac{1}{2^{\alpha-1}}-3\delta^{\alpha}]^{1/\alpha},
		\end{align*} 
		where $\delta=\text{min}\{\delta_1,\delta_2\}$, $\delta_1=\frac{1-p}{4}$
        ~~\text{and}~~$\delta_2= \frac{1+3p}{4}$ are eigenvalues of  $\rho^{wer}_2$ with multiplicity $3$ and $1$, respectively.
        \end{theorem}
        \begin{eqnarray}
		  &&\text{{\bf{Proof:}} Let the Conditional Rényi $\alpha-$entropy CRAE}< 0{}\nonumber\\&&
             \Rightarrow S_{\alpha}(A|B)_{\rho^{wer}_{2}(p)}  < 0 
			\Rightarrow \log_{2}[\frac{\left(\frac{1+3p}{4}\right)^{\alpha}+3 \left(\frac{1-p}{4}\right)^{\alpha}}{2\left(\frac{1}{2}\right)^{\alpha}}]  > 0 {}\nonumber \\&&
			\Rightarrow \left(\frac{1+3p}{4}\right)^{\alpha}+3 \left(\frac{1-p}{4}\right)^{\alpha} > 2\left(\frac{1}{2}\right)^{\alpha}{}\nonumber \\&&			\Rightarrow \text{FEF}(\rho^{wer}_2) >\left[\frac{1}{2^{\alpha-1}}-3\delta^{\alpha}\right]^{1/\alpha}; ~ \alpha>1. {}\nonumber
		    \end{eqnarray}

        \begin{theorem}
            If conditional Rényi $\alpha-$entropy (CRAE) of two-qubit Werner state is negative, then it is useful for teleportation whenever $\delta<\left[\frac{2^\alpha-2^{\alpha-1}}{3.2^{2\alpha-1}}\right]^{\frac{1}{\alpha}}$, where $\delta=\text{min}\{\delta_1,\delta_2\}$, $\delta_1=\frac{1-p}{4}$
        ~~\text{and}~~$\delta_2= \frac{1+3p}{4}$ are eigenvalues of  $\rho^{wer}_2$ with multiplicity $3$ and $1$, respectively.
        \end{theorem}
		\noindent {\bf{Proof:}} For $\delta<\left[\frac{2^\alpha-2^{\alpha-1}}{3.2^{2\alpha-1}}\right]^{\frac{1}{\alpha}} \Rightarrow \left[\frac{1}{2^{\alpha-1}}-3\delta^{\alpha}\right]^{1/\alpha}>\frac{1}{2}$ and by using Theorem (\ref{t2}) we have $\Rightarrow \text{FEF}(\rho^{wer}_2)>\frac{1}{2}$.
		\subsection{Weyl state}
		We consider the two qubit Weyl state, 
		\begin{equation}
			\rho^{weyl}_{2}= \frac{1}{4}[\mathbb{I}_2 \otimes \mathbb{I}_2  +\sum_{i=1}^{3} t_{i}(\sigma_{i} \otimes \sigma_{i})]. \label{e15}
		\end{equation}
		The conditional Rényi 2-entropy of $ \rho^{weyl}_{2}  $ is given by $S_2 (A|B)_{\rho^{weyl}_{2}}=1-\log_{2}(1+t_{1}^2 + t_{2}^2 + t_{3}^2) $ and the fully entangled fraction of Weyl state is given by $\text{FEF}(\rho^{weyl}_2)=\frac{1}{4}[1+N(\rho^{weyl}_2)]$, where $N(\rho^{weyl}_2)=Tr|T|$ and $T$ is correlation matrix.
		\begin{theorem}\label{t4}
			If the conditional Rényi $2-$entropy (CR$2$E) of two-qubit Weyl state (i.e., $\rho^{weyl}_2$) is negative then the fully entangled fraction is strictly greater than half, and hence useful for teleportation i.e.,
		\begin{align*}
			 \text{CR$2$E}(\rho^{weyl}_2) < 0 \Rightarrow FEF(\rho^{weyl}_2)>\frac{1}{2}.
		\end{align*}
		\end{theorem}
       \begin{eqnarray}
           &&\text{{\bf{Proof:}}~Let CR$2$E$(\rho^{weyl}_2)$ is negative}
           \Rightarrow S_{2}(A|B)_{\rho^{weyl}_2}< 0 {}\nonumber\\&&
           \Rightarrow t_{1}^2+t_{2}^2+t_{3}^2  > 1 {}\nonumber \\&&
           \Rightarrow (|t_{1}|+|t_{2}|+|t_{3}|)^2  > 1~\text{where} ~|t_{1}|\neq|t_{2}|\neq|t_{3}|\neq0 {}\nonumber \\&&
           \Rightarrow |t_{1}|+|t_{2}|+|t_{3}|  > 1 
		 \Rightarrow N(\rho^{weyl}_2)  > 1  {}\nonumber \\&&
         \Rightarrow  \text{FEF}(\rho^{weyl}_2)  > \frac{1}{2}. {}\nonumber 
       \end{eqnarray}
		
  \begin{theorem}\label{t5}
			If the fully entangled fraction of two-qubit Weyl state is strictly greater than half then the conditional Rényi $2-$entropy (CR$2$E) is bounded above by $\log_{2}[\frac{1}{1-R}]$ i.e.,
		\begin{align*}
			FEF(\rho^{weyl}_2) >\frac{1}{2} \Rightarrow \text{CR$2$E}(\rho^{weyl}_2) < \log_{2}\left[\frac{1}{1-R}\right],
		\end{align*}
		where $R=|t_{1}||t_{2}|+|t_{1}||t_{3}|+|t_{2}||t_{3}|$ and $0<R<1$.\\
		\end{theorem}
        \begin{eqnarray}
           &&\text{{\bf{Proof:}}~ Let $ \text{FEF}(\rho^{weyl}_2) $ $>\frac{1}{2}$}
           \Rightarrow N(\rho^{weyl}_2)  > 1{}\nonumber\\&&
           \Rightarrow |t_{1}|+|t_{2}|+|t_{3}| > 1 {}\nonumber \\&&
           \Rightarrow 1 + t_{1}^2+t_{2}^2+t_{3}^2 + 2(|t_{1}||t_{2}|+|t_{1}||t_{3}|+|t_{2}||t_{3}|) > 2 {}\nonumber \\&&
           \Rightarrow 1+t_{1}^2+t_{2}^2+t_{3}^2 > 2-2R {}\nonumber \\&&
           \Rightarrow \log_{2}[1+t_{1}^2+t_{2}^2+t_{3}^2]  > \log_{2}[2-2R] {}\nonumber \\&&
		 \Rightarrow 1-\log_{2}(1+t_{1}^2+t_{2}^2+t_{3}^2)  < 1-\log_{2}[2-2R]  {}\nonumber \\&&
         \Rightarrow S_{2}(A|B)_{\rho^{weyl}_2}< \log_{2}\left[\frac{1}{1-R}\right]. {}\nonumber 
       \end{eqnarray}
  Having discussed the relations between FEF and QCE in two qubit systems, we now move to systems consisting of two qudits in the next section. 
		\section{Conditional Entropies and Fully Entangled fraction in $d \otimes d$ system} \label{IV}
  In this section we  extend our study from $2 \otimes 2$ systems to states taken from $d \otimes d$ systems. Here we take the cases of $d \otimes d$ extensions of  Werner State, generalized Bell diagonal state, rank deficient state and isotropic state.
\subsection{Isotropic states}
We consider the isotropic state  \cite{horodecki1999reduction}, which is given by 
		\begin{align}
			\rho_{d}^{iso}=F|\psi_{d}^{+}\rangle\langle\psi_{d}^{+}|+(1-F)\frac{I_{d\times d}-|\psi_{d}^{+}\rangle\langle\psi_{d}^{+}|}{d^{2}-1} ,\label{e16}
		\end{align}
		where $|\psi_{d}^{+}\rangle=\frac{1}{\sqrt{d}}\sum_{i=1}|ii\rangle$ and $ \text{FEF}(\rho_{d}^{iso})= F$. Conditional von Neumann entropy (CVNE) is given by
		    \begin{eqnarray}
			S(A|B)_{\rho^{iso}_{d}}=-F\log_{2}(F)-(1-F)\log_{2}\left(\frac{1-F}{d^2-1}\right)-\log_{2}(d). \label{e17}
		\end{eqnarray}

		\subsubsection{{\textbf{Relation with conditional von Neumann entropy}}}
		\begin{theorem}\label{t6}
			If conditional von Neumann entropy (CVNE) of isotropic state is negative then fully entangled fraction of isotropic state is bounded below by $\log_{(d^{2}-1)}[\frac{d^2-1}{d}]$ i.e.,
		\begin{align*}
			\text{CVNE}(\rho_{d}^{iso}) < 0 \Rightarrow F>\log_{(d^{2}-1)}\left[\frac{d^2-1}{d}\right].
		\end{align*}
		\end{theorem}
        \begin{eqnarray}
           &&\text{{\bf{Proof:}}~ Let CVNE is negative}
           \Rightarrow S(A|B)_{\rho^{iso}_{d}} < 0{}\nonumber\\&&
           \Rightarrow -F\log_{2}(F)-(1-F)\log_{2}\left(\frac{1-F}{d^2-1}\right)-\log_{2}(d) < 0 {}\nonumber \\&&
           \Rightarrow (1-F)\log_{2}(d^{2}-1)  < \log_{2}(d) - H(F,1-F){}\nonumber\\&&
			\Rightarrow \log_{2}(d^{2}-1)- F\log_{2}(d^{2}-1)  < \log_{2}(d){}\nonumber\\&&
			\Rightarrow F\log_{2}(d^{2}-1)  > \log_{2}\left(\frac{d^{2}-1}{d}\right){}\nonumber\\&&
	       \Rightarrow F  > \log_{(d^{2}-1)}\left[\frac{d^2-1}{d}\right].{}\nonumber
       \end{eqnarray}
		Since, $\log_{(d^{2}-1)}[\frac{d^2-1}{d}]>\frac{1}{d}$ for all $d>2$, and for $d=2$, negative CVNE implies that $F>\frac{1}{2}$. Thus, we have the following corollary.
		\begin{corollary}\label{c6.1}
			If conditional von Neumann entropy (CVNE) of isotropic state is negative then fully entangled fraction of isotropic state is strictly greater than $\frac{1}{d}$ i.e., \text{CVNE}$(\rho_{d}^{iso}) < 0 \Rightarrow F>\frac{1}{d}.$
		\end{corollary}
		\subsubsection{{\textbf{Relation with conditional Rényi 2-entropy}}}
        Conditional Renyi 2-entropy (CR$2$E) of isotropic state is given by 
		\begin{align}
			S_{2}(A|B)_{\rho^{iso}_{d}}=\log_{2}\left[\frac{d^2-1}{d(d^2-1)F^2+d(1-F)^2}\right]. \label{e18}
		\end{align} 
		\begin{theorem}\label{t7}
			The conditional Renyi 2-entropy (CR$2$E) of isotropic state with fully entangled fraction $F$ in $d \otimes d$ system is negative iff fully entangled fraction is lower bounded by $\frac{1+\sqrt{1+d(d^{2}-d-1)}}{d^2}$ i.e.,
		\begin{align*}
			\text{CR$2$E}(\rho_{d}^{iso}) < 0 \Leftrightarrow F  > \frac{1+\sqrt{1+d(d^{2}-d-1)}}{d^2}.
		\end{align*}
		\end{theorem}
  \begin{eqnarray}
            &&\text{{\bf{Proof:}} Let CR$2$E is negative}\Leftrightarrow S_{2}(A|B)_{\rho^{iso}_{d}}<0 {}\nonumber\\&&
		\Leftrightarrow \log_{2}\left[\frac{d^2-1}{d(d^2-1)F^2+d(1-F)^2}\right]<0{}\nonumber\\&& 
			\Leftrightarrow \log_{2}\left[\frac{d^2-1}{d(d^2-1)F^2+d(1-F)^2}\right]<\log_{2}(1) {}\nonumber\\&&
   \Leftrightarrow \frac{d^2-1}{d(d^2-1)F^2+d(1-F)^2}<1{}\nonumber\\&&
			\Leftrightarrow d^{3}F^2 - 2dF + (1-d^2 +d) > 0 {}\nonumber\\&&
		\Leftrightarrow F > \frac{1+\sqrt{1+d(d^{2}-d-1)}}{d^2}.{}\nonumber
        \end{eqnarray}
	\subsection{Werner state in two qudits}
        The Werner state in $d\otimes d$ system is defined as \cite{werner1989quantum}
        \begin{align}
            \rho_{w} = \frac{d-x}{d^3-d} I \otimes I + \frac{dx-1}{d^3-d}V~,~~~~~ x\in[-1,1],
        \end{align}
        where $V$ is swap operator such that $V(|ij\rangle)=|ji\rangle.$ FEF for Werner states is given in \cite{Zhao_2010}, and denoted by $F$. \\
        If $d$ is even, FEF is given by
        \begin{align*}
        F=
           \begin{cases}
              \frac{1+x}{d(d+1)}, & \ \frac{1} 
               {d}\leq x \leq 1 \\ \\
	        \frac{1-x}{d(d-1)}, &  \ -1\leq x<    \frac{1}{d}.
            \end{cases}
         \end{align*}
        If $d$ is odd, FEF is given by 
         \begin{align*}
        F=
           \begin{cases}
              \frac{1+x}{d(d+1)}, & \ \frac{1} 
               {d}\leq x \leq 1 \\ \\
	        \frac{d^{2}-d^{2}x+dx+d-2}{d^2(d^2-1)}, &  \ -1\leq x<    \frac{1}{d}
            \end{cases}
         \end{align*}
         \subsubsection{{\textbf{Relation with conditional von Neumann entropy}}}
         Eigenvalues of $\rho_{w}$ are $\frac{1+x}{d^2+d}$ and $\frac{1-x}{d^2-d}$ with multiplicity $\frac{d^2+d}{2}$ and $\frac{d^2-d}{2}$, respectively \cite{li2021information}. Conditional von Neumann entropy is given by
            \begin{eqnarray}
            S(A|B) = -\frac{1+x}{2}\log_{2}\frac{1+x}{d^2+d}-\frac{1-x}{2}\log_{2}\frac{1-x}{d^2-d}-\log_{2}d.
        \end{eqnarray}
        
        \noindent  {\bf {Case-I :} For $d$ is even or odd and $\frac{1}{d}\leq x < 1$.}\\
        \begin{theorem}{\label{t8}}
            If $d$ is even or odd and $\frac{1}{d}\leq x<1$, then $F^{\gamma}=4^{-S(A|B)}~\Gamma $ where $\gamma=1+x$ and $\Gamma=\left(\frac{d^2-d}{1-x}\right)^{1-x}\frac{1}{d^2}$.
        \end{theorem}
        \begin{eqnarray}
            &&\text{{\bf{Proof:}}~If $d$ is even or odd and $\frac{1}{d}\leq x<1$} \Rightarrow F=\frac{1+x}{d(d+1)} {}\nonumber\\&&
            S(A|B)=  -\frac{1+x}{2}\log_{2}\frac{1+x}{d^2+d} - \frac{1-x}{2}\log_{2}\frac{1-x}{d^2-d}-\log_{2}d {}\nonumber\\&&
            \Rightarrow S(A|B)=  -\frac{1+x}{2}\log_{2}F - \frac{1-x}{2}\log_{2}\frac{1-x}{d^2-d}-\log_{2}d {}\nonumber\\&&
            \Rightarrow F^{1+x} = 4^{-S(A|B)}\left(\frac{d^2-d}{1-x}\right)^{1-x}\frac{1}{d^2}{\nonumber}
        \end{eqnarray}
        {\bf{Note:}} If $x=1$ then $F=\frac{2^{-S(A|B)}}{d}$.\\
        
       \noindent {\bf{Case-II} For $d$ is even and $-1< x < \frac{1}{d}$.}\\
       \begin{theorem}{\label{t9}}
           If $d$ is even and $-1< x<\frac{1}{d}$, then $F^{\gamma}=4^{-S(A|B)}~\Gamma $ where $\gamma=1-x$ and $\Gamma=\left(\frac{d^2+d}{1+x}\right)^{1+x}\frac{1}{d^2}$.
        \end{theorem}
        \begin{eqnarray}
            &&\text{{\bf{Proof:}}~If $d$ is even and $-1< x<\frac{1}{d}$} \Rightarrow F=\frac{1-x}{d(d-1)} {}\nonumber\\&&
            S(A|B)=  -\frac{1+x}{2}\log_{2}\frac{1+x}{d^2+d} - \frac{1-x}{2}\log_{2}\frac{1-x}{d^2-d}-\log_{2}d {}\nonumber\\&&
            \Rightarrow S(A|B)=  -\frac{1+x}{2}\log_{2}\frac{1+x}{d^2+d} - \frac{1-x}{2}\log_{2}F-\log_{2}d {}\nonumber\\&&
            \Rightarrow F^{1-x} = 4^{-S(A|B)}\left(\frac{d^2+d}{1+x}\right)^{1+x}\frac{1}{d^2}{\nonumber}
        \end{eqnarray}
        {\bf{Note:}} If $x=-1$ then $F=\frac{2^{-S(A|B)}}{d}$.\\
         \subsubsection{{\textbf{Relation with conditional Rényi $\alpha$-entropy}}}
         Conditional Rényi $\alpha$- entropy (CRAE) of $\rho_{w}$ is given by
         \begin{eqnarray}
            S_{\alpha}(A|B)=&&\frac{1}{1-\alpha}\log_{2}[d^{\alpha-1}[\left(\frac{d^2+d}{2}\right)\left(\frac{1+x}{d^2+d}\right)^{\alpha} +\left(\frac{d^2-d}{2}\right)\left(\frac{1-x}{d^2-d}\right)^{\alpha}]],~~~~ \alpha > 1.
        \end{eqnarray}
 \noindent {\bf{Case-I} For $d$ is even or odd and $\frac{1}{d}\leq x \leq 1$}
        \begin{theorem}{\label{t12}}
            If $d$ is even or odd and $\frac{1}{d}\leq x\leq1$, then $F^{\alpha}=\Delta $ where $\Delta=\frac{2.2^{(1-\alpha)S_{\alpha}(A|B)}-d^{\alpha-1}(d^2-d)^{1-\alpha}(1-x)^{\alpha}}{d^{\alpha-1}(d^2+d)}$.
        \end{theorem}
        \begin{eqnarray}
            &&\text{{\bf{Proof:}}~If $d$ is even or odd and $\frac{1}{d}\leq x\leq1$} \Rightarrow F=\frac{1+x}{d(d+1)} {}\nonumber\\&&
             S_{\alpha}(A|B)=\frac{1}{1-\alpha}\log_{2}[d^{\alpha-1}[\left(\frac{d^2+d}{2}\right)\left(\frac{1+x}{d^2+d}\right)^{\alpha}+\left(\frac{d^2-d}{2}\right)\left(\frac{1-x}{d^2-d}\right)^{\alpha}]]{}\nonumber\\&&
             \Rightarrow 2^{(1-\alpha)S_{\alpha}(A|B)}= \frac{d^{\alpha-1}(d^2+d)F^{\alpha}+d^{\alpha-1}(d^2-d)^{1-\alpha}(1-x)^{\alpha}}{2}{}\nonumber\\&&
             \Rightarrow F^{\alpha} =\frac{2.2^{(1-\alpha)S_{\alpha}(A|B)}-d^{\alpha-1}(d^2-d)^{1-\alpha}(1-x)^{\alpha}}{d^{\alpha-1}(d^2+d)}.{}\nonumber 
        \end{eqnarray}

     \noindent {\bf{Case-II} For $d$ is even and $-1\leq x < \frac{1}{d}.$}\\
        \begin{theorem}
            If $d$ is even and $-1\leq x < \frac{1}{d}$, then $F^{\alpha}=\Delta $ where $\Delta=\frac{2.2^{(1-\alpha)S_{\alpha}(A|B)}-d^{\alpha-1}(d^2+d)^{1-\alpha}(1+x)^{\alpha}}{d^{\alpha-1}(d^2-d)}$.
        \end{theorem}
        \begin{eqnarray}
            &&\text{{\bf{Proof:}}~If $d$ is even and $-1\leq x < \frac{1}{d}$} \Rightarrow F=\frac{1-x}{d(d-1)} {}\nonumber\\&&
             S_{\alpha}(A|B)=\frac{1}{1-\alpha}\log_{2}[d^{\alpha-1}[\left(\frac{d^2+d}{2}\right)\left(\frac{1+x}{d^2+d}\right)^{\alpha}+\left(\frac{d^2-d}{2}\right)\left(\frac{1-x}{d^2-d}\right)^{\alpha}]]{}\nonumber\\&&
             \Rightarrow 2^{(1-\alpha)S_{\alpha}(A|B)}= \frac{d^{\alpha-1}(d^2+d)^{1-\alpha}(1+x)^{\alpha}+d^{\alpha-1}(d^2-d)F^{\alpha}}{2}{}\nonumber\\&&
             \Rightarrow F^{\alpha} =\frac{2.2^{(1-\alpha)S_{\alpha}(A|B)}-d^{\alpha-1}(d^2+d)^{1-\alpha}(1+x)^{\alpha}}{d^{\alpha-1}(d^2-d)}.{}\nonumber 
        \end{eqnarray}
        \subsection{Rank-deficient states}
         Rank-two entangled state in $d\otimes d$ system, denoted by $\rho_{rd}$, is given by \cite{horodecki1999general, verstraete2003optimal}
         \begin{eqnarray}
             \rho_{rd}= p|\psi_{d}^{+}\rangle\langle\psi_{d}^{+}|+(1-p)|01\rangle\langle 01|, 
             ~~p\in(0,1].
         \end{eqnarray}
         where $|\psi_{d}^{+}\rangle=\frac{1}{\sqrt{d}}\sum_{i=1}|ii\rangle$. If $d\geq 4$, state $\rho_{rd}$ is useful for teleportation for all $p\in(0,1]$, and when $d\leq3$ state $\rho_{rd}$ is useful for teleportation for $p\in(\frac{1}{d},1]$ \cite{li2021activating}. We discuss here the relation between conditional entropies and fully entangled fraction of the rank deficient state.
         \subsubsection{{\textbf{Relation with conditional von Neumann entropy}}}
         CVNE of $\rho_{rd}$ is given by 
             \begin{eqnarray}
             S(A|B)_{\rho_{rd}}=\frac{dp-p}{d}\log_{2}\left(\frac{p}{d}\right)+\frac{d-dp+p}{d}\log_{2}\left(\frac{d-dp+p}{d}\right) + H(p),
         \end{eqnarray}
         where $H(p)$ is binary entropy i.e., $H(p)=-p\log_{2}(p)-(1-p)\log_{2}(1-p)$. Table \ref{T1} notes the negativity of CVNE and is usefulness for teleportation for various dimensions of state, i.e., $d=2,3,4,5,6$.
        \begin{center}
            \begin{table}
            \caption{Negativity of CVNE}\label{T1}
            \begin{tabular}{|c|c|c|}
                \hline
                 Dimension$(d)$ & Useful for teleportation  & Negative CVNE\\
                \hline
                $2$ & $0.5 < p\leq 1 $ & $0.666667< p<1$\\
                \hline
                $3$ & $0.3333 < p\leq 1$  & $0.241217 < p<1$\\
                \hline 
                $4$ & $0 < p \leq 1 $ & $0.0409511 < p<1$\\
                \hline
                $5$ & $0 < p \leq 1 $ &  $0.00433229 < p<1$ \\
                \hline
                $6$ & $0 < p \leq 1 $ &  $0.000349461 < p<1$\\
                \hline
           \end{tabular}
           \end{table}
        \end{center}
         Thus, we can say that if the CVNE of the rank-deficient state is negative then it is useful for teleportation for the dimension $d=2,4,5,6$. But for $d=3$, the scenario is different, there is a range of parameter values in which CVNE is negative but still it is not useful for teleportation.
        \subsubsection{{\textbf{Relation with conditional Rényi 2-entropy}}}
        Now, we check the relation between conditional Renyi 2-entropy and fully entangled fraction of the state $\rho_{rd}$. Conditional Renyi 2-entropy is given by 
        \begin{equation}
            S_{2}(A|B)_{\rho_{rd}}=\log_{2}\left[\frac{(d-1)p^2 +(d-dp+p)^2}{d^2(p^2 +(1-p)^2)}\right].
        \end{equation}
        We observe that CR$2$E is negative if $p > \frac{2}{d+1}$. For $d\leq 3$, state $\rho_{rd}$ is useful for teleportation when $p > \frac{1}{d}$. Now, we consider a function $f(d)=\frac{2}{d+1}-\frac{1}{d}$ and find that $f(d)>0 $ for $d\geq 2$. So, if CR$2$E is negative, then it is useful for teleportation whenever $d\leq 3$. Since, state $\rho_{rd}$ is useful for teleportation for all $d\geq4$ when $p\in (0,1]$, so if CR$2$E is negative, then it is useful for teleportation for $d\geq 4$. Thus, we can conclude that if CR$2$E is negative then state $\rho_{rd}$ is useful for teleportation.
        \subsection{Generalized Bell Diagonal States}
        Consider a maximally entangled pure state on Hilbert space $\mathbb{H}_{AB}=\mathbb{H}_{A}\otimes \mathbb{H}_{B}$, $\text{dim}~ \mathbb{H}_{A} = \text{dim}~ \mathbb{H}_{B} = d$, is given by $|\psi_{d}^{+}\rangle=\frac{1}{\sqrt{d}}\sum_{i=0}^{d^2-1}|ii\rangle$ 
        Here, $|\psi_{d}^{+}\rangle$ is also known as the Bell state. Now, we consider unitary operators, also known as Weyl operators, defined by 
        \begin{equation}
            U_{mn}|i\rangle = \eta^{m(i-n)}|i-n\rangle,  
        \end{equation}
        where $\eta= e^{\frac{2\pi j}{d}}, m, n, i = 0,1,2,..., d-1$.\\
        The action of the unitary operator $U_{mn}$ on one of the subsystems of the chosen Bell state produces other mutually orthogonal Bell states and is given by $|\psi_{mn}\rangle = (I\otimes U_{mn})|\psi_{d}^{+}\rangle$. The Bell projectors corresponding to $|\psi_{mn}\rangle$ is given by 
        \begin{eqnarray}
            P_{mn}&& = |\psi_{mn}\rangle\langle\psi_{mn}|{}\nonumber\\&&=(I\otimes U_{mn})|\psi_{d}^{+}\rangle\langle\psi_{d}^{+}|(I\otimes U_{mn}^{\dagger}).
        \end{eqnarray}
        For our convenience, we change the denotation of $\{P_{mn}\}_{m,n=0}^{d-1}$ to $\{P_i\}_{i=0}^{d^2-1}$. We now consider the simplex of Bell projectors \cite{bertlmann2008bound,bertlmann2008geometric,baumgartner2006state,baumgartner2007special,baumgartner2008geometry,chruscinski2010class} given by $\rho_{gb} = \sum_{i=0}^{d^2-1}p_{i}P_{i}~,~~~~ p_{i} \geq 0,~~~~ \sum_{i=0}^{d^2-1}p_{i}=1$. 
        The fully entangled fraction from Equation(\ref{e7a}) of $\rho$ is max$\{p_{i}\}_{i=0}^{d^2-1}$, where $\{p_{i}\}_{i=0}^{d^2-1}$ are the eigenvalues of state $\rho_{gb}$ and $i=0,1,...,d^2-1.$ So, FEF($\rho_{gb}$) is $F=$ max$\{p_i\}_{i=0}^{d^2-1}$. In the subsequent subsection, we investigate the relation between fully entangled fractions and conditional entropies.
        \subsubsection{{\textbf{Relation with conditional von Neumann entropy}}}
        Conditional von Neumann entropy (CVNE) of $\rho_{gb}$ is given by $S(A|B)=-\sum_{i=0}^{d^2-1}p_{i}\log_2(p_{i})-\log_{2}(d)$.
        \begin{theorem}{\label{t16}}
            If conditional von Neumann entropy of state $\rho_{gb}$ is negative, then $F^{F}d\beta > 1$ where $F$ is fully entangled fraction, $d$ is dimension, and $\beta=\prod_{i=0}^{d^2-2}p_{i}^{p_{i}}$ i.e., product of all $p_{i}^{p_i}$ except the term for largest $p_{i}$, $p_{i}>0$.
        \end{theorem}
        \begin{eqnarray}
           &&\text{{\bf{Proof:}}~ Let conditional von Neumann entropy is negative,}{}\nonumber\\&&
           \Rightarrow S(A|B)   < 0 
            \Rightarrow S(AB)  < \log_{2}(d) {}\nonumber\\&&
            \Rightarrow  -\sum_{i=0}^{d^2-1}p_{i}\log_{2}(p_{i}) < \log_{2}(d) {}\nonumber\\&&
            \Rightarrow -\left(F\log_{2}(F)+\sum_{i=0}^{d^2-2}p_{i}\log_{2}(p_{i})\right)  < \log_{2}(d) {}\nonumber\\&&
            \Rightarrow -F\log_{2}(F) < \log_{2}(d) + \sum_{i=0}^{d^2-2}p_{i}\log_{2}(p_{i})  {}\nonumber\\&&
            \Rightarrow -F\log_{2}(F)< \log_{2}(d) + \log_{2}\left(\prod_{i=0}^{d^2-2}p_{i}^{p_{i}}\right) {}\nonumber\\&&
            \Rightarrow F^{F}d\beta > 1.{}\nonumber
       \end{eqnarray}
       \begin{theorem}{\label{gbdst1}}
            If $F>\frac{1}{d}$ then $S(A|B)<Y$ where $Y=\log_{2}\left(\frac{d^{F-1}}{\prod_{i=0}^{d^2-2}p_{i}^{p_{i}}}\right)$, and $\prod_{i=0}^{d^2-2}p_{i}^{p_{i}}$ is product of all $p_{i}^{p_i}$ except the term for largest $p_{i}$, $p_{i}>0$.
            \end{theorem}
         \begin{eqnarray}
         &&\text{{\bf{Proof:}}~ Here we have }F > \frac{1}{d}
           \Rightarrow -F\log_{2}(F)< F\log_{2}(d) {}\nonumber\\&&
       S(A|B)=-\sum_{i=0}^{d^2-1}p_{i}\log_{2}(p_{i})- \log_{2}(d){}\nonumber\\&&
                  =-F\log_{2}(F)-\sum_{i=0}^{d^2-2}p_{i}\log_{2}(p_{i})-\log_{2}(d){}\nonumber\\&&
                  <F\log_{2}(d)-\sum_{i=0}^{d^2-2}p_{i}\log_{2}(p_{i})-\log_{2}(d){}\nonumber\\&&
                  < \log_{2}(d^{F-1}) - \log_{2}(\prod_{i=0}^{d^2-2}p_{i}^{p_{i}}){}\nonumber\\&&
                  < \log_{2}(\frac{d^{F-1}}{\prod_{i=0}^{d^2-2}p_{i}^{p_{i}}})~,~~p_{i}>0{}\nonumber
        \end{eqnarray}
        \subsubsection{\textbf{Relation with other $ \alpha- $ conditional entropies}}
        Conditional Rényi $\alpha$- entropy (CRAE) of $\rho_{gb}$ is given by
            $S_{\alpha}(A|B)=\frac{1}{1-\alpha}\log_{2}\left(d^{\alpha-1}\sum_{i=0}^{d^2-1}p_{i}^{\alpha}\right) ~,~~~ \alpha>1.$
        \begin{theorem}
            If conditional Rényi $\alpha$- entropy (CRAE) of $\rho_{gb}$ is negative then $F^{\alpha} > \left(\frac{1-d^{\alpha-1}X}{d^{\alpha-1}}\right)$ where $X=\sum_{i=0}^{d^2-2}p_{i}^{\alpha}$ i.e., sum of all $p_{i}^\alpha$ except the term for largest $p_i$.
            \end{theorem}
        \begin{eqnarray}
&&\text{{\bf{Proof:}}~ Let the conditional Rényi $\alpha$- entropy (CRAE) of}{}\nonumber\\&&
           \text{ $\rho_{gb}$ is negative.}
          \Rightarrow S_{\alpha}(A|B)  < 0 \nonumber\\&&
            \Rightarrow \frac{1}{1-\alpha}\log_{2}\left(d^{\alpha-1}\sum_{i=0}^{d^2-1}p_{i}^{\alpha}\right) < 0 \nonumber\\&&
            \Rightarrow \log_{2}\left(d^{\alpha-1}\sum_{i=0}^{d^2-1}p_{i}^{\alpha}\right) > 0
            \Rightarrow d^{\alpha-1}\sum_{i=0}^{d^2-1}p_{i}^{\alpha} > 1 \nonumber\\&&
            \Rightarrow F^{\alpha} + \sum_{i=0}^{d^2-2}p_{i}^{\alpha} > \frac{1}{d^{\alpha-1}}
            \Rightarrow F^{\alpha} > \frac{1}{d^{\alpha-1}} - \sum_{i=0}^{d^2-2}p_{i}^{\alpha}.\nonumber
       \end{eqnarray}
        \begin{theorem} \label{gbdst2}
            If $F>\frac{1}{d}$ then $S_{\alpha}(A|B)< \frac{1}{1-\alpha}\log_{2}\left(\frac{1+d^{\alpha}Z}{d}\right)$ where $Z=\sum_{i=0}^{d^2-2}p_{i}^{\alpha}$ i.e., sum of all $p_{i}^\alpha$ except the term for largest $p_i$. 
        \end{theorem}
        \begin{eqnarray}
&&\text{{\bf{Proof:}}~ Let $F  > \frac{1}{d}$}
          \Rightarrow F^{\alpha}  > \frac{1}{d^\alpha} 
            \Rightarrow F^{\alpha}+Z > Z + \frac{1}{d^\alpha}{}\nonumber\\&&
            \Rightarrow \frac{1}{1-\alpha}\log_{2}\left(F^{\alpha}+Z\right) < \frac{1}{1-\alpha}\log_{2}\left(Z + \frac{1}{d^\alpha}\right) {}\nonumber\\&&
            \Rightarrow \frac{1}{1-\alpha}\log_{2}\left(\sum_{i=0}^{d^2-1}p_{i}^{\alpha}\right)<\frac{1}{1-\alpha}\log_{2}\left(\frac{1+d^{\alpha}Z}{d^\alpha}\right){}\nonumber\\&&
            \Rightarrow \frac{1}{1-\alpha}\log_{2}\left(\sum_{i=0}^{d^2-1}p_{i}^{\alpha}\right)-\frac{1}{1-\alpha}\log_{2}\left(\frac{1}{d^{\alpha-1}}\right){}\nonumber\\&&<\frac{1}{1-\alpha}\log_{2}\left(\frac{1+d^{\alpha}Z}{d^\alpha}\right)-\frac{1}{1-\alpha}\log_{2}\left(\frac{1}{d^{\alpha-1}}\right){}\nonumber\\&&
            \Rightarrow S_{\alpha}(A|B) < \frac{1}{1-\alpha}\log_{2}\left(\frac{1+d^{\alpha}Z}{d}\right).{}\nonumber
       \end{eqnarray}

       As noted earlier, there is an intricate relation of FEF of quantum states and its nonlocal characteristics when several copies of the state is taken. Having obtained relations between FEF and QCE for several classes of quantum states, we now proceed to investigate the relation between $k-$ copy nonlocality (both steering and standard Bell nonlocality) and QCE in the subsequent sections. 
        
		\section{K-copy steerability and conditional entropies}\label{V}
		In this section we investigate the relation of conditional entropies and $k-$copy steerability for low dimension and few copies of isotropic state. 
  \subsection{Isotropic state}
		\subsubsection{{\textbf{Relation with conditional von Neumann entropy}}}
		Conditional von Neumann entropy of isotropic state $\rho^{iso}_{d}$ is given by Equation (\ref{e17}).\\
        
\noindent  {\bf {Case-I} For low dimension $d=2$.}\\

\noindent Conditional von Neumann entropy (i.e., CVNE) of isotropic state for low dimension (i.e., $d=2$) is negative for $F>0.81071$. In \cite{quintino2016superactivation}, it is mentioned that any state is $k-$copy steerable for projective measurement if 
		\begin{align}
			F^k > \frac{(1+d^k)(\sum_{n=1}^{d^k}\frac{1}{n}-1)-d^k}{d^{2k}}. \label{e19}
		\end{align}
		From \cite{quintino2016superactivation}, we observe that $k=7$ is the minimum number of copies $k$ such that two-qubit isotropic state $\rho^{iso}_{2}$ is $k-$copy steerable, and hence we obtained the range for $F$ i.e., $F>0.600034$. Thus, for $k\geq 7$, we can conclude that if CVNE of two-qubit isotropic state is negative then it is $k-$copy steerable.\\
  
\noindent {\bf{Case-II} For few copies $k=2$.}\\
        
\noindent  Here, we discuss the $k-$copy steering with few copies of state. From \cite{quintino2016superactivation}, for $k=2$, we know that $\rho^{iso}_{d}$ is $k-$copy steerable for projective measurement whenever $d\geq6$. Here, we check the relation of CVNE and $k-$copy steering for $d=6$. We observe that CVNE is negative for $F>0.673671$, and it is $k-$copy steerable whenever $F>0.257221$. Thus, for dimension six, we can say that if CVNE is negative, then it is $k-$copy steerable.

		\subsubsection{{\textbf{Relation with conditional Rényi 2-entropy}}}
		We see here the relation between conditional Renyi 2-entropy(CR$2$E) and $k-$copy steerability of isotropic states. CR$2$E is given by Equation (\ref{e18}).\\
  
        \noindent {\bf{Case-I} For low dimension $d=2$ .}\\
        
        CR$2$E of isotropic state for low dimension $d=2$ is negative for $F>0.683013$. From \cite{quintino2016superactivation}, we observe that $k=7$ is the minimum number of copies $k$ such that two-qubit isotropic state $\rho^{iso}_{2}$ is $k-$copy steerable, and hence we obtained the range for $F$ i.e., $F>0.600034$. Thus, for $k\geq 7$, we can conclude that if CR$2$E of two-qubit isotropic state is negative then it is $k-$copy steerable.\\
        
        \noindent {\bf{Case-II} For few copies $k=2$.}\\ 
        
        Here, we discuss the $k-$copy steering with few copies of state. From \cite{quintino2016superactivation}, for $k=2$, we know that $\rho^{iso}_{d}$ is $k-$copy steerable for projective measurement whenever $d\geq6$. Here, we check the relation of CR$2$E and $k-$copy steering for $d=6$. We observe that CR$2$E is negative for $F>0.395243$, and it is $k-$copy steerable whenever $F>0.257221$. Thus, for dimension six, we can say that if CR$2$E is negative, then it is $k-$copy steerable.
		\section{k-copy nonlocality and conditional entropies}\label{VI}
		
	\subsection{Weyl states}
	By using Theorem (\ref{t4}), we know that two-qubit Weyl states with negative CR$2$E are useful for teleportation i.e., $\text{FEF}(\rho^{weyl}_2)>\frac{1}{2}$, and from \cite{cavalcanti2013all} we know that all quantum states useful for teleportation are nonlocal resources and hence $k-$copy nonlocal. Thus, we have the following result.\\
 
\begin{observation} \label{weyl-cr2e}
All two-qubit Weyl states with negative CR$2$E are $k-$copy nonlocal and hence, are nonlocal resources i.e.,
		\begin{align*}
			\text{CR$2$E}(\rho^{weyl}_2)<0 \Rightarrow \text{k-copy nonlocal.} 
		\end{align*}
 \end{observation} 
	\subsection{Non-Weyl state}
	Consider a two-qubit non-Weyl state given by \cite{bowles2016sufficient,cavalcanti2013all}
	\begin{align}
		\rho_2^{nl}=p|\psi_{x}\rangle\langle\psi_{x}| +(1-p)[\rho_{x}^{A}\otimes \frac{\mathbb{I}}{2}], \label{e20}
	\end{align} 
	where $|\psi_{x}\rangle = \cos x|00\rangle +\sin x|11\rangle $, $\rho_{x}^{A}=\text{Tr}_B(|\psi_x\rangle\langle\psi_{x}|)$; $0<x\leq\frac{\pi}{4}$, $0\leq p\leq 1.$ In \cite{cavalcanti2013all}, if $\text{FEF}(\rho_2^{nl})>\frac{1}{2}$ then $p\geq \frac{1}{1+2\sin 2x}$. Hence, $\rho_2^{nl}$ is $k-$copy nonlocal for all $0<x\leq\frac{\pi}{4}$ when $p\geq \frac{1}{1+2\sin 2x}$. CR$2$E is negative when $p>\frac{1}{\sqrt{3}}$ for all $0<x\leq\frac{\pi}{4}$. Now, we consider a function $f(x)=\frac{1}{\sqrt{3}}-\frac{1}{1+2\sin 2x}$ for all $0<x\leq\frac{\pi}{4}$. We observe that function $f(x)$ is positive for $\frac{1}{2}\text{arcsin}(\frac{\sqrt{3}-1}{2})<x\leq\frac{\pi}{4}$. Thus, we can conclude that if CR$2$E is negative then it is $k-$copy nonlocal.
	\subsection{Noisy state}
	We consider a pure state $|\psi\rangle$ with its Schmidt decomposition given by 
	\begin{align}
		|\psi\rangle=\sum_{i=0}^{d-1}\lambda_{i}|ii\rangle, \label{e21}
	\end{align}
	where $\lambda_{i}$ are real, strictly positive and satisfies normalised condition $\sum_{i=0}^{d-1}\lambda_{i}^{2}=1$. In \cite{almeida2007noise}, the mixture of an arbitrary pure state $|\psi\rangle$ with maximally mixed noise is given by 
	\begin{align}
		\delta_{d}^{ns}= p|\psi\rangle\langle\psi| + (1-p)\frac{I_{d}}{d}\otimes\frac{I_{d}}{d}~. \label{e22}
	\end{align}
	In \cite{cavalcanti2013all}, it is given that the state $\delta_{d}^{ns}$ is $k-$copy nonlocal if 
	\begin{align}
		p > \frac{d-1}{d(\sum_{i=0}^{d-1}\lambda_{i})^2-1}~. \label{e23}
	\end{align}
	Conditional Rényi 2-entropy (CR$2$E) of $\delta_{d}^{ns}$ is negative if 
	\begin{align}
		p > \frac{\sqrt{d-1}}{\sqrt{d-1 + d^{2}(1-\sum_{i=0}^{d-1}\lambda_{i}^{4})}}~.\label{e24}
	\end{align}
	In the below, we see the relation between $k-$copy nonlocal and negative conditional entropies for $d=3,4,5$. Here, we are fixing the Schmidt coefficient of state $\delta_{d}^{ns}$ in Equation (\ref{e21}) for respective dimension. Considering the Schmidt coefficient ($\lambda_{0}=\frac{1}{\sqrt{2}},\lambda_{1}=\frac{1}{\sqrt{3}},\lambda_{2}=\frac{1}{\sqrt{6}}$), ($\lambda_{0}=\lambda_{1}=\frac{1}{\sqrt{3}},\lambda_{2}=\lambda_{3}=\frac{1}{\sqrt{6}}$) and ($\lambda_{0}=\frac{1}{\sqrt{3}},\lambda_{1}=\frac{1}{\sqrt{7}},\lambda_{2}=\sqrt{\frac{2}{21}}, \lambda_{3}=\sqrt{\frac{4}{21}}, \lambda_{4}=\sqrt{\frac{5}{21}}$~) 
	for dimension $d=3, 4$ and $5$, respectively. We observe that the state is $k-$copy nonlocal and have negative CR$2$E in the range ($p>0.263305$ \& $p>0.516398$), ($p>0.206292$ \& $p>0.45399$) and ($p>0.174342$ \& $p>0.415577$) for dimension $d=3,4$ and $5$, respectively. Thus, we can say that if the state $\delta_{d}^{ns}(d=3,4,5)$ has negative CR$2$E for the above restriction, then it is $k-$copy nonlocal. Also, we observe that the state is $k-$copy nonlocal and have negative CVNE in the range ($p>0.263305$ \& $p>0.728901$), ($p>0.206292$ \& $p>0.699086$) and ($p>0.174342$ \& $p>0.685898$) for dimension $d=3,4$ and $5$, respectively. Thus, we can say that if the state $\delta_{d}^{ns}(d=3,4,5)$ has negative CVNE for the above restriction, then it is $k-$copy nonlocal.

	
\section{Application}\label{VII}

Having seen some implications of the relations in the context of multi-copy nonlocality, we now proceed to provide some applications in this section.

\subsection{Work Extraction}
Consider any quantum state $\rho \in  \mathfrak{B}(\mathbb{H}_{AB}) $ where $\mathbb{H}_{AB}=\mathbb{H}_{A}\otimes \mathbb{H}_{B}$, $\text{dim}~ \mathbb{H}_{A} = \text{dim}~ \mathbb{H}_{B} = d$. In \cite{hsieh2017work}, the authors introduced a framework and defined minimal deterministic average work cost, denoted by $W_{g}(\rho)$, in the IID (independent and identically distributed) limit of the erasure process according to \cite{rio2011thermodynamic}. The authors in \cite{hsieh2017work} used the twirling operation and FEF of isotropic states to achieve bounds to  $W_{g}(\rho)$ in terms of FEF. We obtain bounds for  $W_{g}(\rho)$, using the relations obtained between FEF and QCE obtained in our work, for some special class of states detailed below. 

Bounds for $W_{g}(\rho)$ is given by \cite{rio2011thermodynamic,hsieh2017work}
\begin{align}
    -W_{g}(\rho)\leq S(A|B)_{\rho}kT\ln2~, \label{efw}
\end{align}
where $k$ is the Boltzmann constant, $k>>1$ \cite{hsieh2017work}, and $T$ is environmental temperature. 
\subsubsection{\textbf{Werner state in two qubits}}
Here, we investigate the work gain for the two-qubit Werner state. By Theorem(\ref{t1}), we have bounds of conditional von Neumann entropy of the two-qubit Werner state. Using Theorem(\ref{t1}) and Equation(\ref{efw}), we obtain the following results for the two-qubit Werner state.\\
\begin{observation} \label{work-werner}
If fully entangled fraction of two-qubit Werner states is strictly greater than half i.e., $FEF(\rho^{wer}_2)>\frac{1}{2}$, then 
    \begin{align*}
        W_{g}(\rho^{wer}_2) > kT3\delta\ln(2\delta)~,
    \end{align*}
   where $\delta=\text{min}\{\delta_1,\delta_2\}$, $\delta_1=\frac{1-p}{4}$~~\text{and}~~$\delta_2= \frac{1+3p}{4}$ are eigenvalues of  $\rho^{wer}_2$ with multiplicity $3$ and $1$, respectively.
\end{observation}
\subsubsection{\textbf{Generalized Bell Diagonal States}}
Here, we investigate the work gain for the generalized Bell diagonal state $\rho$. By Theorem(\ref{gbdst1}), we have bounds of conditional von Neumann entropy of generalized Bell diagonal state. Using Theorem(\ref{gbdst1}) and Equation(\ref{efw}), we obtain the following results for the generalized Bell diagonal state.\\
\begin{observation}   \label{work-genbell} 
If fully entangled fraction of generalized Bell diagonal states is strictly greater than $\frac{1}{d}$ i.e., $FEF(\rho)>\frac{1}{d}$, then 
    \begin{align*}
        W_{g}(\rho) > kT\ln(Q)~,
    \end{align*}
    where $Q=\left(\frac{\prod_{i=0}^{d^2-2}p_{i}^{p_{i}}}{d^{F-1}}\right).$
\end{observation}

\subsection{Faithful entanglement}
Detection of entanglement in quantum systems is a core problem in quantum information science. Negativity of conditional entropy is a signature of entanglement. Within the class of entangled states there are certain states which can be detected by a proximity-measure based witness. Such proximity-measured witnesses are usually constructed from FEF. In this subsection, we show that such entangled states can also be detected through conditional entropies, which is a consequence of the relations obtained between FEF and CE. 

Consider any quantum state $\rho \in  \mathfrak{B}(\mathbb{H}_{AB}) $ where $\mathbb{H}_{AB}=\mathbb{H}_{A}\otimes \mathbb{H}_{B}$, $\text{dim}~ \mathbb{H}_{A} = \text{dim}~ \mathbb{H}_{B} = d$. Now, consider a pure entangled state $|\psi\rangle$ such that the fidelity of $\rho$ with respect to pure entangled state $|\psi\rangle$ is denoted by $F_{|\psi\rangle}(\rho)$, and is given as $F_{|\psi\rangle}(\rho)=\langle\psi|\rho|\psi\rangle$. Fidelity is a measure of proximity of a state $\rho$ with a pure entangled state $|\psi\rangle$. The fidelity-based entanglement witness can be given as $W_{|\psi\rangle} = \eta I-|\psi\rangle\langle\psi|$ for some suitably chosen real number $\eta$. On measuring the observable $W_{|\psi\rangle}$, the quantity $\textbf{Tr}(\rho W_{|\psi\rangle})$ is obtained. If $F_{|\psi\rangle}(\rho)$ is greater than some threshold value $\eta$, then the witness operator detects the presence of entanglement in $\rho$. Such entangled states are termed as \textit{faithful} \cite{weilenmann2020entanglement,riccardi2021exploring,guhne2021geometry}. If the state $|\psi\rangle$ is a maximally entangled pure state i.e., $|\psi\rangle = |\phi^{+}\rangle=\frac{1}{\sqrt{d}}\sum_{i=1}^{d}|ii\rangle$, then the smallest possible value of $\eta$ is $\frac{1}{d}$, and the witness operator is given as $W_{|\psi\rangle} = \frac{I}{d}-|\psi\rangle\langle\psi|$. From \cite{guhne2021geometry}, we know that a state $\rho$ is faithful if and only if $\text{FEF}(\rho)>\frac{1}{d}$. Since, faithful entanglement is directly related to FEF, we can now characterize faithful entanglement in terms of conditional entropies. Since, direct evaluation of FEF is computationally difficult albeit for some special states, the conditions obtained in terms of conditional entropies can simplify the characterization. 
\subsubsection{\textbf{Weyl states in two qubits}}
Here, we investigate the faithful entanglement of the two-qubit Weyl state. Using Theorem(\ref{t4}), and the result from \cite{guhne2021geometry}, we obtain the following results for the two-qubit Weyl state.
\begin{observation} \label{faithful-weyl1}
    If the conditional Rényi $2-$entropy (CR$2$E) of two-qubit Weyl state (i.e., $\rho^{weyl}_2$) is negative, then $\rho^{weyl}_2$ is faithful entangled state.
\end{observation}
Using Theorem(\ref{t5}) and the result from \cite{guhne2021geometry}, we obtain the following results for the two-qubit Weyl state.
\begin{observation} \label{faithful-weyl2}
    If two-qubit Weyl state is faithful entangled then the conditional Rényi $2-$entropy (CR$2$E) is bounded above by $\log_{2}[\frac{1}{1-R}]$ where $R=|t_{1}||t_{2}|+|t_{1}||t_{3}|+|t_{2}||t_{3}|$ and $0<R<1$.
\end{observation}
\subsubsection{\textbf{Isotropic state in two qudits}}
Here, we investigate the faithful entanglement of two-qudit isotropic states. Using Corollary(\ref{c6.1}) and result from \cite{guhne2021geometry}, we obtain the following result.
\begin{observation} \label{faithful-iso}
    If conditional von Neumann entropy (CVNE) of two-qudit isotropic state is negative then it is faithful entangled.
\end{observation}
From Theorem(\ref{t7}), we have 
\begin{align*}
	\text{CR$2$E}(\rho_{d}^{iso}) < 0 \Leftrightarrow F  > \frac{1+\sqrt{1+d(d^{2}-d-1)}}{d^2}.
\end{align*}
By doing some algebraic calculations, we obtain that $\frac{1+\sqrt{1+d(d^{2}-d-1)}}{d^2}>\frac{1}{d}$ for all $d>1$, and hence $F>\frac{1}{d}$. So, we conclude the following
\begin{align}
    \text{CR$2$E}(\rho_{d}^{iso}) < 0 \Rightarrow F  > \frac{1}{d}. \label{apl-iso}
\end{align}
Now, using Equation(\ref{apl-iso}) and the result from \cite{guhne2021geometry}, we obtain the following result for two-qudit isotropic state.
\begin{observation}
    If conditional Renyi 2-entropy (CR$2$E) of isotropic state with fully entangled fraction $F$ in $d \otimes d$ system is negative then the state is faithful.
\end{observation}
\subsubsection{\textbf{Generalized Bell Diagonal States}}
Here, we investigate the faithful entanglement for the generalized Bell diagonal state $\rho$.
Using Theorem(\ref{gbdst1}) and the result from \cite{guhne2021geometry}, we see the relation between faithful entanglement and von Neumann entropy and obtain the following results for generalized Bell diagonal state.

\begin{observation}
    If the generalized Bell diagonal state $\rho$ is faithful then $S(A|B)<Y$ where $Y=\log_{2}\left(\frac{d^{F-1}}{\prod_{i=0}^{d^2-2}p_{i}^{p_{i}}}\right), p_{i}>0$.
\end{observation}

Now, using Theorem(\ref{gbdst2}) and the result from \cite{guhne2021geometry}, we see the relation between faithful entanglement and the conditional Rényi $\alpha$- entropy (CRAE), and obtain the following results for generalized Bell diagonal state.
\begin{observation}
    If the generalized Bell diagonal state $\rho$ is faithful then $S_{\alpha}(A|B)< \frac{1}{1-\alpha}\log_{2}\left(\frac{1+d^{\alpha}Z}{d}\right)$ where $Z=\sum_{i=0}^{d^2-2}p_{i}^{\alpha}$ i.e., sum of all $p_{i}^\alpha$ except the term for largest $p_i$. 
\end{observation}

\subsection{Entropic uncertainty relation}
Consider two incompatible measurements $X$ and $Y$ such that $H(X)$ and $H(Y)$ are Shannon entropies corresponding to the outcomes of measurements when $X$ and $Y$ are performed on a quantum state. Then the entropic uncertainty relation \cite{robertson1929uncertainty} is given by \cite{kraus1987complementary,maassen1988generalized}
\begin{align}
    H(X) + H(Y) \geq\log_{2}\frac{1}{c}, \label{eup1}
\end{align}
where $c=\text{max}_{i,j}|\langle x_{i}|y_{j}\rangle|^2$ such that $|x_{i}\rangle$ and $|y_{j}\rangle$ are eigenvector of $X$ and $Y$, respectively.

The operational interpretation of the above Equation(\ref{eup1}) can be understood through the so-called guessing game \cite{berta2010uncertainty}. However, the uncertainty relation above is no longer valid when one of the parties has access to quantum memory. The entropic uncertainty relation in the quantum memory scenario is stated as \cite{berta2010uncertainty}
\begin{align}
    H(X|B) + H(Y|B) \geq\log_{2}\frac{1}{c} + S(A|B), \label{eup2}
\end{align}
where $H(X|B)$ or $H(Y|B)$ denote the amount of uncertainty of measurement $X$ or $Y$ given the information stored in quantum memory $B$. Here, from Equation(\ref{eup2}), we observe that if $S(A|B)$ (i.e., CVNE) is negative, then the state prepared by Bob overcomes the original uncertainty bound of Equation(\ref{eup1}). 

Now, we investigate the entropic uncertainty relation for the two-qudit Werner state. By using Theorem(\ref{t8}) and the Equation(\ref{eup2}), we obtain the following observations.
\begin{observation}\label{ob10}
    If $d$ is even or odd and $\frac{1}{d}\leq x<1$, then 
    \begin{align*}
         H(X|B) + H(Y|B) \geq\log_{2}\frac{1}{c} + \frac{1}{2}\log_{2}(\frac{\Gamma}{F^{\gamma}})
    \end{align*}
    where $\gamma=1+x$ and $\Gamma=\left(\frac{d^2-d}{1-x}\right)^{1-x}\frac{1}{d^2}$.
\end{observation}
Now, if we consider $F\leq\frac{1}{d}$ in Observation(\ref{ob10}), then we obtain the following lower bounds
\begin{align*}
     H(X|B) + H(Y|B) \geq\log_{2}\frac{1}{c} + \frac{1}{2}\log_{2}(\Gamma d^{\gamma})
\end{align*} 
where $\gamma=1+x$ and $\Gamma=\left(\frac{d^2-d}{1-x}\right)^{1-x}\frac{1}{d^2}$.
In a similar way, by using Theorem(\ref{t9}) and the Equation(\ref{eup2}), we obtain the following observation for the two-qudit Werner state.
\begin{observation} \label{ob11}
    If $d$ is even and $-1< x<\frac{1}{d}$, then \begin{align*}
         H(X|B) + H(Y|B) \geq\log_{2}\frac{1}{c} + \frac{1}{2}\log_{2}(\frac{\Gamma}{F^{\gamma}})
    \end{align*}
    where $\gamma=1-x$ and $\Gamma=\left(\frac{d^2+d}{1+x}\right)^{1+x}\frac{1}{d^2}$.
\end{observation}

In Observation(\ref{ob11}), if we consider $F\leq\frac{1}{d}$, then we obtain the following lower bounds
\begin{align*}
     H(X|B) + H(Y|B) \geq\log_{2}\frac{1}{c} + \frac{1}{2}\log_{2}(\Gamma d^{\gamma})
\end{align*} 
where $\gamma=1-x$ and $\Gamma=\left(\frac{d^2+d}{1+x}\right)^{1+x}\frac{1}{d^2}$.

In both observations we are also able to find the lower bounds of the entropic uncertainty relation for the states in the class of two-qudit Werner states that are not useful for teleportation.

        \section{Conclusion}\label{VIII}
        The present work discusses the relationship between conditional entropies and fully entangled fraction, as they provide important yardsticks for the merit of quantum states in information processing tasks. In $2 \otimes 2$ system, for Werner state, we obtain the upper bounds of conditional von Neumann entropy provided its fully entangled fraction is greater than half. Also, the lower bound of the fully entangled fraction is obtained in terms of eigenvalue if its conditional Rényi $\alpha-$entropy is negative. For a two-qubit Weyl state, the negativity of conditional Rényi $2-$entropy is a sufficient condition for being useful for teleportation, and the upper bounds to the conditional Rényi $2-$entropy is obtained whenever it is useful for teleportation. Further, extending our studies to arbitrary $d \otimes d$ systems, we observe that the fully entangled fraction of an isotropic state is lower bounded provided its conditional von Neumann entropy is negative. We study the relation between fully entangled fraction and conditional entropies for for two qudit Werner states, rank deficient states and generalized Bell diagonal states. Also, we investigate the relation of $k-$copy steerability, $k-$copy nonlocality and conditional entropies. If the conditional von Neumann entropy of two-qubit isotropic is negative then it is $k-$copy steerable, same behaviour holds in the case of conditional Rényi $2-$entropy. On considering a few copies ($k=2$) of isotropic state in six dimensions, the negativity of conditional von Neumann entropy and conditional Rényi $2-$entropy guarantees the $k-$copy steerability. The negativity of conditional Rényi $2-$entropy of two-qubit Weyl state gives the sufficient condition for $k-$copy nonlocality. The same holds for some non-Weyl states. We show that the negativity of conditional Rényi $2-$entropy of state (mixture of arbitrary pure state and maximally mixed noise) for dimension $d=3,4,5$ is the sufficient condition for $k-$copy nonlocality, and the same holds whenever the conditional von Neumann entropy is negative. Further, we discuss the implications of the relations obtained in the context of work extraction, faithful entanglement, and entropic uncertainty relation.

        We observe that in general the relation between FEF (particularly when the threshold value $\text{FEF}>1/d$) and negativity of conditional entropy is intricate. This calls for further probe for general quantum states. As we observe, such relations are significantly linked with quantum information processing protocols and thus further investigation will shed light on the efficacy of quantum states in information processing tasks. The implications of this study also carry significance in foundational studies like $k-$ copy steerability and $k-$ copy nonlocality and thus calls for further attention.\\ 
        
\noindent{\bf{Acknowledgement:}} NG acknowledges support from SERB-DST (India) MATRICS grant vide file number MTR/2022/000101.

\section*{Declaration}
\noindent All the authors contributed equally to the manuscript.\\
{\bf{Data availability:}} Data sharing was not applicable to this article as no data sets were generated or analyzed during the current study.\\
{\bf{Competing interests:}} The authors have no competing interests to declare. All co-authors have seen and agree with the contents of the manuscript, and there is no financial interest to report.

  \bibliographystyle{apsrev4-2-titles.bst}
  \bibliography{ref}

	\end{document}